\documentclass{PoS}
\pdfoutput=1

\usepackage{amsmath,amssymb,amsthm,amsfonts}
\usepackage{MnSymbol}
\usepackage{mathrsfs}
\usepackage{slashed}
\usepackage{graphicx}
\usepackage{subfigure}
\usepackage{color,rotating}
\usepackage{dsfont}
\usepackage{setspace}
\usepackage{verbatim}
\usepackage{fancyhdr}

\newcommand{\beq}{\begin{equation}}
\newcommand{\eeq}{\end{equation}}
\newcommand{\beqa}{\begin{eqnarray}}
\newcommand{\eeqa}{\end{eqnarray}}
\newcommand{\bfc}{\begin{figure}[h!]\begin{center}}
\newcommand{\efc}{\end{center}\end{figure}}

\def\Fig#1{Fig.~\ref{#1}}

\newcommand{\bea}{\begin{eqnarray}}
\newcommand{\eea}{\end{eqnarray}}
\newcommand{\Eq}[1]{Eq.~(\ref{#1})}

\def\eq#1{(\ref{#1})}
\def\Eq#1{Eq.~(\ref{#1})}
\newcommand {\apgt} {\ {\raise-.5ex\hbox{$\buildrel>\over\sim$}}\ }
\newcommand {\aplt} {\ {\raise-.5ex\hbox{$\buildrel<\over\sim$}}\ }

\def\s0#1#2{\mbox{\small{$ \frac{#1}{#2} $}}}
\def\0#1#2{\frac{#1}{#2}}

\def\Cz{{\mathcal Z}}

\def\CP{{\mathcal P}}

\def\CV{{\mathcal V}}

\newcommand{\tr}{\mathrm{tr}}
\newcommand{\I}{\mathrm{i}}
\newcommand{\be}{\begin{eqnarray}}
\newcommand{\ee}{\end{eqnarray}}

\newcommand{\Nc}{N_{\rm{c}}}

\title{Quark Confinement from Correlation Functions} \ShortTitle{Confinement
  from Correlation Functions}
\author{\speaker{Leonard Fister}\\
  Department of Mathematical Physics, National University of Ireland Maynooth, Maynooth Co. Kildare, Ireland, and\\
  Institut f\"ur Theoretische Physik, Universit\"at Heidelberg, Philosophenweg 16, 69120 Heidelberg, Germany.\\
  E-mail: \email{leonard@thphys.nuim.ie}}
\author{Jan M. Pawlowski\\
  Institut f\"ur Theoretische Physik, Universit\"at Heidelberg,
  Philosophenweg 16, 69120 Heidelberg, Germany, and\\
  ExtreMe Matter Institute EMMI, GSI, Planckstr. 1,
  D-64291 Darmstadt, Germany.\\
  E-mail: \email{j.pawlowski@thphys.uni-heidelberg.de}} \abstract{ We 
  study quark confinement by computing the Polyakov loop potential in
  Yang--Mills theory within different non-perturbative functional
  continuum approaches \cite{Fister:2013bh}. We extend previous studies in the formalism of
  the functional renormalisation group and complement those with
  findings from Dyson--Schwinger equations and
  two-particle-irreducible functionals. These methods are formulated
  in terms of low order Green functions. This allows to identify a
  criterion for confinement solely in terms of the low-momentum
  behaviour of correlators.}
\FullConference{Xth Quark Confinement and the Hadron Spectrum,\\
  October 8-12, 2012\\
  TUM Campus Garching, Munich, Germany}

\begin{document}
\section{Introduction}
In recent years much progress has been made in our understanding of
the strongly-correlated low energy regime of QCD in terms of
gauge-fixed correlation functions. This progress is tightly linked to
the advance in our understanding of the basic phenomena of low energy
QCD, strong chiral symmetry breaking and confinement. Confinement
denotes the absence of colour charged asymptotic states. For
infinitely heavy quarks this can be related to the free energy of a
single quark $F_{q}$: Bringing a single quark into the system would
require an infinite amount of energy in the confined phase, whereas a
finite energy would suffice in the deconfined phase. The operator
which creates a static single quark is related to the Polyakov loop
$L$,
\beq\label{eq:Polloop}
L = \frac{1}{\Nc} \tr_{\rm f}\CP\, e^{\I g \int_0^\beta dx_0\,
    {A}_0(x_0,x)} \,, \quad {\rm and } \ \langle L \rangle \sim {\rm exp}^{-F_{q}/T} \,,
\eeq
with the trace in the fundamental representation of $SU(N_c)$ of the
closed Wilson line in time direction. Here, $\CP$ denotes path
ordering, $A$ is the gauge field and the inverse temperature
$\beta=1/T$.  This links the confinement-deconfinement phase
transition in the Yang--Mills system to the order-disorder phase
transition of center symmetry $Z_N$ in $SU(N_c)$: Under center
transformations $z\in Z_N$ the Polyakov loop transforms with $L\to z\,
L$ in the fundamental representation. In the center-symmetric,
confining phase the order parameter vanishes, $\langle L\rangle=0$, while in the
center-broken, deconfined phase it is finite, $\langle L\rangle\neq 0$.

It has been shown that $L[\langle A_0\rangle]$ also constitutes an
order parameter for static quark confinement
\cite{Braun:2007bx,Marhauser:2008fz}, as $L[\langle A_0\rangle] \geq
\langle L[A_0] \rangle$ in the deconfined but strictly $L[\langle
A_0\rangle]=0$ in the confined phase.  However, $L[\langle
A_0\rangle]$ is easily accessible within functional continuum methods,
as $\langle A_0\rangle$ can be straightforwardly computed from the
equation of motion, that is via the full effective potential
$V[A_0]$. It was shown within the functional renormalisation group
(FRG) approach \cite{Braun:2007bx,
  Marhauser:2008fz,Braun:2009gm,Pawlowski:2010ht}, that $V[A_0]$ has a
closed representation in terms of the full propagators of the gluons,
ghosts (and quarks) in constant $A_0$-backgrounds.  With this
connection a confinement criterion in terms of the infrared behaviour
of the propagators was derived \cite{Braun:2007bx}.

By now $\langle L[A_0] \rangle$ and the effective Polyakov loop
potential $V[A_0]$ has been computed in Yang--Mills theory in the
Landau gauge \cite{Fister:2013bh,Braun:2007bx,Braun:2010cy} and in the Polyakov
gauge \cite{Marhauser:2008fz}. The potential obtained in different
gauges agree well which provides a check of the gauge
independence. Recently, these studies have been extended to the
Coulomb gauge \cite{Reinhardt:2012qe}. More recently, lattice results
for the potential have been obtained
\cite{Diakonov:2012dx,Greensite:2012dy}. In \cite{Braun:2009gm} the
Yang--Mills studies have been extended to fully dynamical two-flavour
QCD. In \cite{Fukushima:2012qa,Kashiwa:2012td}, $V[A_0]$ is used in an
effective model approach to QCD with interesting applications to
thermodynamic observables.

Here we report on our work in \cite{Fister:2013bh}, in which we compute
the Polyakov loop effective potential $V[A_0]$ in the background field
formalism in Landau--deWitt gauge within different functional
approaches, \cite{Fister:2013bh}. In Section~\ref{sec:funPolPot} we
derive representations for $V[A_0]$ with the FRG, Dyson--Schwinger
equations (DSEs) and two-particle irreducible (2PI) functionals. In
Section~\ref{sec:confcrit} we deduce a confinement criterion in terms
of the propagators: Infrared suppression of gluons but non-suppression
of ghosts confines static quarks at sufficiently low temperatures. In
Section~\ref{sec:resultsPolPot} numerical results for $V[A_0]$ are
presented, which are based on the finite temperature propagators
computed within the FRG in \cite{Fister:2011uw,Fister:2011um,
  Fister:Diss}.

\section{Polyakov Loop Potential from Functional Methods}\label{sec:funPolPot}
Most functional approaches are based on closed expressions for the
effective action or derivatives thereof in terms of full correlation
functions. Hence, the knowledge of the latter in constant
$A_0$-backgrounds allows us to compute the Polyakov loop potential
$V[A_0]$. In turn, confinement requires $V[A_0]$ to have minima at the
confining values for $A_0$, i.e.\ in $SU(\Nc)$ the center-symmetric
points.  With $\left(\bar{c}\right) c$ being the (anti-)ghosts, in the
background formalism the Polyakov loop potential is the
gauge-invariant effective action, $\Gamma[A_0;\phi]$ with
$\phi=(a=A-A_0,c,\bar c)$, evaluted on a constant $A_0$-background and
vanishing fluctuation fields, $V[A_0^{\rm const}]:=\Gamma[A^{\rm
  const}_0;0]/\left(\beta \rm \CV\right)$. $\CV$ is the
three-dimensional spatial volume. The Polyakov loop, \eq{eq:Polloop},
is then computed at the minimum $\langle A_0\rangle:=A_{0,\rm min}$.
Functional equations for the effective action depend on the
correlation functions of fluctuation fields only. By gauge covariance
the correlation functions in the background Landau gauge, 
$\Gamma^{(n)}[\bar A]$, are directly related to those in the Landau gauge, 
$\Gamma^{(n)}[0]$. Neglecting subleading corrections, for more details see 
\cite{Fister:2013bh}, the two-point functions of gluon and the ghost on
constant backgrounds are schematically given by 
\beq
\Gamma_a^{(2)}[A;0]= \sum_{L/T} \left(-D^2\right) Z_{L/T} P^{L/T}\ ,\
{\rm and} \quad \Gamma_c^{(2)}[A;0]= \left(-D^2\right) Z_{c}\,, 
\eeq
where the $P^{L/T}$ are the projection operators for the
chromoelectric/chromomagnetic modes, and
$Z_{L/T/c}(-D_0^2,-\vec{D}^2)$ are the wave-function renormalisations
of the corresponding gluon and ghost modes evaluated on covariant
momenta $D$.
\begin{figure}[t]
\begin{center}
  \subfigure[Flow equation for the Polyakov loop
  potential.]{\includegraphics[width=.42\textwidth]{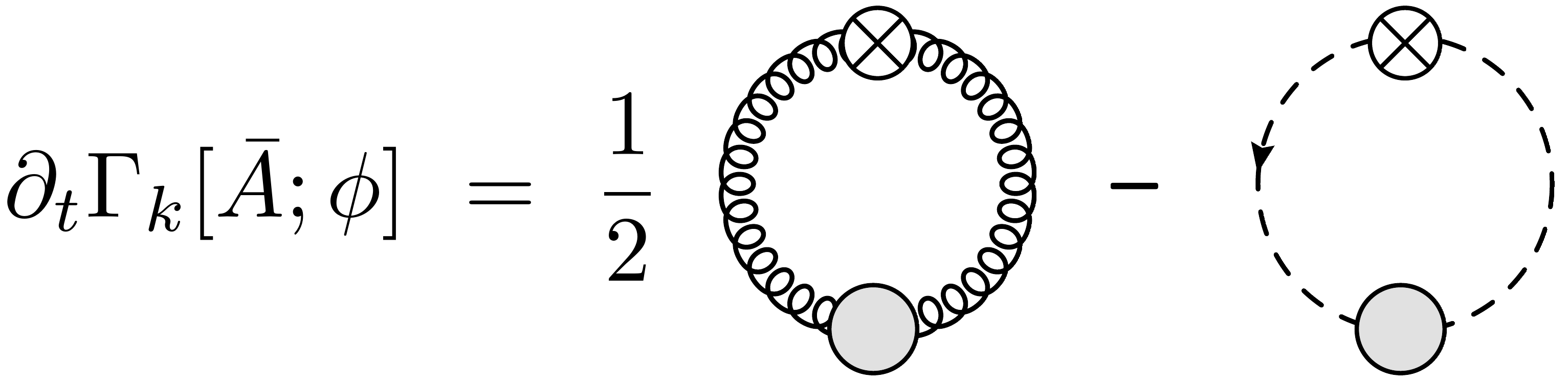}\label{fig:funflow}}\hfill
  \subfigure[DSE for the background gluon one-point
  function.]{\includegraphics[width=.54\textwidth]{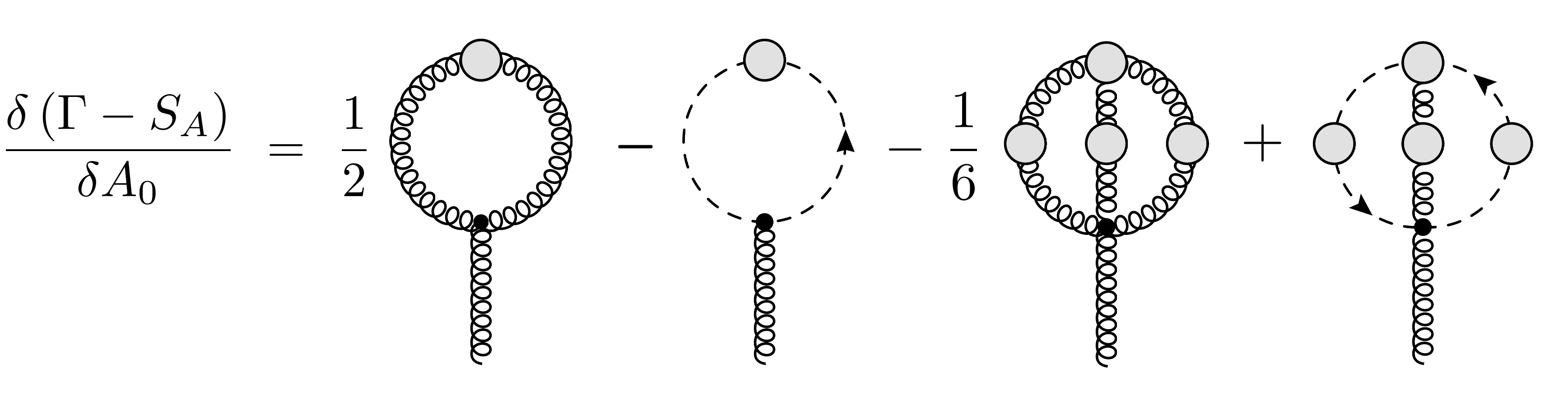}\label{fig:DSE}}\hfill
\caption[]{Functional equations for the (derivative of) the Polyakov
  loop potential. The gluons are represented by wiggly, the ghosts by
  dashed lines. Filled circles indicate full correlators. Crossed
  circles denote the regulator insertion $\partial_t R_k$.}
\label{fig:funeqs}
\end{center}
\end{figure}
%
The FRG representation of the Polyakov loop potential has been used in
\cite{Braun:2007bx, Marhauser:2008fz, Braun:2010cy}. The exact flow equation for the
effective action, $\Gamma_k[\bar A;\phi]$, is given solely in terms of the propagators,
\beqa 
  \partial_t \Gamma_{k}[\bar A;\phi] =\! 
  \sumint_p\left( \frac{1}{2}  \left(G_a\right)^{ab}_{\mu\nu}[\bar A;\phi](p,p)
  \, {\partial_t} \left(R_a\right)_{\nu\mu}^{ba}(p)-
 \left(G_{c}\right)^{ab}[\bar A;\phi](p,p)
  \, {\partial_t} \left(R_{c}\right)^{ba}(p)\right)\,,
\label{eq:funflow}
\eeqa 
where the integration and Matsubara summation involve the respective
modes. Further, $t=\ln k$, with $k$ being a infrared cut-off
scale controlled by the regulator $R_{a/c}$. At $k$ the propagators are
\beq
G_{a/c,k}(p)=\left(\left(\Gamma_{a/c}\right)_k (p)+ R_k(p)\right)^{-1}\,.
\eeq
The diagrammatic representation of the flow equation is given in \Fig{fig:funflow}. 

Finally, by integration we get the Polyakov loop potential
\begin{equation}\label{eq:flowV}
V[A_0]=V_{k=\Lambda}[A_0]+\0{1}{\beta \rm \CV} \int_\Lambda^0 dt\,
\partial_t \Gamma_k[A_0;0]\,,
\end{equation}
where $V'_{k=\Lambda}[A_0]\to 0$ for $\Lambda/T\gg 1$.  The DSE for
Yang--Mills theory relevant for the Polyakov loop potential is that
originating in a derivative of the effective action $\Gamma$ w.r.t.\
the $A_0$-background at fixed fluctuation. It can be written in terms
of full propagators and vertices and the renormalised classical
action,
\beqa 
\hspace{-.3cm}   \0{\delta(\Gamma[A_0;0]-S_A[A_0;0])}{\delta A_0(x)}=
\012 S^{(3)}_{A_0 aa} G_{a} -
S^{(3)}_{A_0 c\bar c} G_{c}
 -\016 S^{(4)}_{A_0 aaa} G_{a}^3 \Gamma^{(3)}_{aaa} + 
S^{(4)}_{A_0  a c\bar c } G_{c}^2 G_{a} \Gamma^{(3)}_{ac \bar c}\,,
\label{eq:DSE}
\eeqa 
with $S^{(n)}_{\phi_1\ldots\phi_n}$ and
$\Gamma^{(n)}_{\phi_1\ldots\phi_n}$ denoting derivatives of the
renormalised classical action $S_A$ and effective action with respect
to the fields $\phi_i$. For constant backgrounds the DSE equation for
Polyakov loop potential is given by
\begin{equation}\label{eq:DSEPolPot}
\0{\partial V[A_0]}{\partial A_0}=
\0{1}{\beta\CV}  \0{\partial \Gamma[A_0;0]}{\partial A_0}\,,
\end{equation}
with \eq{eq:DSE} for the right hand side of \eq{eq:DSEPolPot}. In
comparison with the DSE for fluctuating fields only the vertices
differ. This also leads to the emergence of the two-loop diagram
involving ghosts. Diagrammatically the DSE is given in \Fig{fig:DSE}.

For details on the different representations see \cite{Fister:2013bh}. There 
it is also shown that for the present purpose the 2PI representation of the Polyakov
loop potential is identical to the DSE \eq{eq:DSE}.

\section{Confinement Criterion}\label{sec:confcrit}
Here we discuss the restriction of the infrared behaviour of the
propagators that follows from the FRG. Here, all loops in the full
flow, \eq{eq:funflow}, effectively boil down to expressions of the
form
\begin{equation}\label{eq:genloop}
 \sim 4\pi T \sum_{n\in \mathbb{Z}}\int_0^\infty\!\! \0{d |\vec{p\,}|}{(2 \pi)^3} \
\vec{p\,}^2 \frac{\partial_t R_k(x)}{ x\, Z(-D_0^2, \vec{p\,}^2)+
\Delta m^2( D,A_0)+R_k(x)}\,,
\end{equation}
where $x=-D_0^2+\vec{p\,}^2$ and $Z=Z_L,Z_T,Z_c$. In \eq{eq:genloop}
we do not include the overall minus sign of the ghost
contribution. The function $R_k$ stands for the scalar parts of the
regulators, which are chosen to be $Z$-independent and
$O(4)$-symmetric, i.e.\ $R_k(x)= x\, r(x/k^2)$.  The terms $\Delta
m^2(D, 0)=0\,,$ with $\Delta m^2 =\Delta m^2_L,\Delta m^2_T,\Delta
m^2_c$, are thermal corrections to the Polyakov loop, which disappear
with powers of the temperature for $T\to 0$, and for the present
purpose we will neglect these terms, \cite{Fister:2013bh},
\beq
\Delta m^2 (D,A_0) \equiv 0\,.
\label{eq:Deltam0}
\eeq
At vanishing temperature we regain $O(4)$-symmetry and $Z(-D_0^2,
\vec{p\,}^2)\to Z(-D_0^2+\vec {p\,}^2)$. In terms of scaling
coefficients $\kappa$ at vanishing temperature,
\begin{equation}\label{eq:scalingcoeffs}
Z(x)=\Cz\, x^\kappa\,, 
\end{equation}
the corresponding propagators exhibit non-thermal mass gaps $m_{\rm
  gap}$ for $\kappa<0$. Computing the expression \eq{eq:genloop}
leads to a deconfining potential in $A_0$: Taking a derivative w.r.t.\
$A_0$ it vanishes at $A_0=0$ and the center-symmetric points, i.e. where
the Polyakov loop vanishes, $L=0$. The second derivative is positive
at $A_0=0$ while it is negative at the center-symmetric points.

Hence, with \eq{eq:Deltam0} the gluon loops in the flow equation
\eq{eq:funflow} give deconfining potentials for the gluons.  In
contrast, the ghosts give a confining contribution due to the relative
minus sign. Thus, confinement in covariant gauges requires the
suppression of gluonic contributions. Indeed, the same mechanism is at
work in the DSE representation \cite{Fister:2013bh}. With the assumption
\eq{eq:Deltam0} this yields a necessary condition for confinement, 
\begin{equation}\label{eq:IRlimits} 
  \lim_{x\to 0}\frac{1}{Z_{L/T}(x)}=0\, \ {\rm and }\ 
 \lim_{x\to 0}\frac{1}{Z_c(x)} > 0\,,  \quad {\rm thus, }\quad
 \kappa_{L/T} < 0\, \ {\rm and }\  \ \kappa_c\geq 0\,,
\end{equation}
where $\kappa_{L/T}$ are the anomalous dimensions of the
chromoelectric and chromomagnetic gluon.  In \eq{eq:IRlimits} the
condition involving the ghost guarantees that its contribution
dominates the trivial gauge mode.  \Eq{eq:IRlimits} is a sufficient
condition for confinement in Yang--Mills propagators (with the
assumption \eq{eq:Deltam0}) as it leads to a vanishing order
parameter, the Polyakov loop expectation value. Note also that it has
been shown \cite{Fischer:2006vf,Fischer:2009tn} within a scaling
analysis that $\kappa_c\geq 0$ has to hold in the infrared.

The infrared limits \eq{eq:IRlimits} are satisfied for both decoupling
type, i.e.\ $\kappa_c=0$ and $\kappa_a=-1$, and scaling type, i.e.\
$\kappa_c=\kappa=-\kappa_a/2$ with $\kappa\approx 0.595$, solutions. 
For an overview of the different solutions see e.g.\
\cite{Fischer:2008uz}. As a consequence, both types exhibit
confinement. Furthermore, they do not affect the critical physics at
the phase transition, as it is triggered in the mid-momentum regime
where both types of solutions agree on a quantitative level.

The confinement criterion provided here a  leads to simple counting
scheme for general theories and gauges, e.g.\ Yang--Mills--Higgs
systems \cite{Fister:2010ah, Fister:2010yw, Alkofer:2010tq, Macher:2011ys}, or QCD with fundamental or adjoint quarks in the Landau gauge,
\cite{Braun:2009gm,Pawlowski:2010ht,Braun:2011fw, Braun:2012zq}. The
respective scenarios are detailed in \cite{Fister:2013bh}.

\section{Results for the Polyakov Loop Potential and $T_c$}\label{sec:resultsPolPot}
\begin{figure}[t]
\begin{center}
\subfigure[Potential for $SU(2)$.]{\includegraphics[width=.48\textwidth]{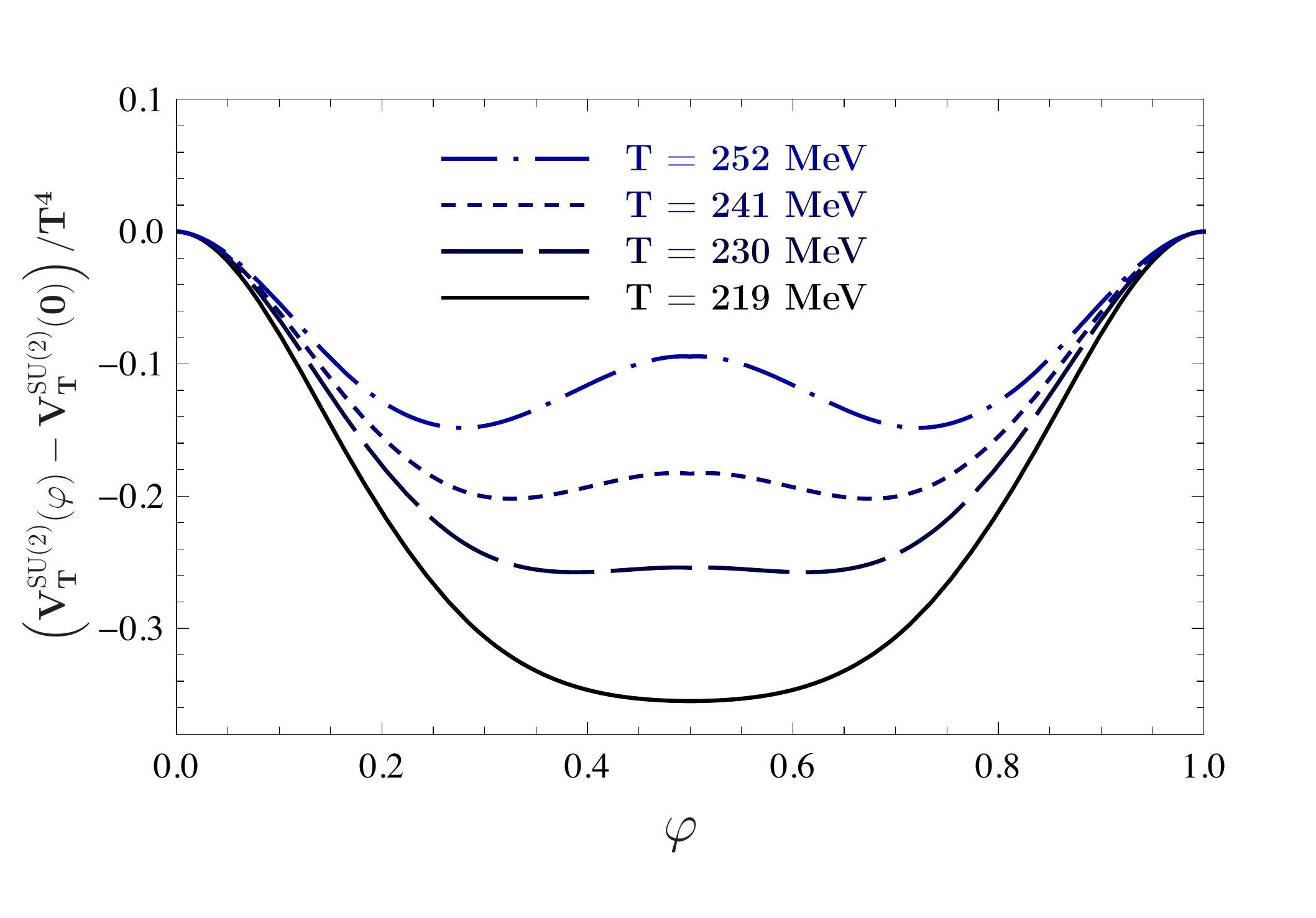}\label{fig:SU2_FRG_T}}\hfill
\subfigure[Potential for $SU(3)$.]{\includegraphics[width=.48\textwidth]{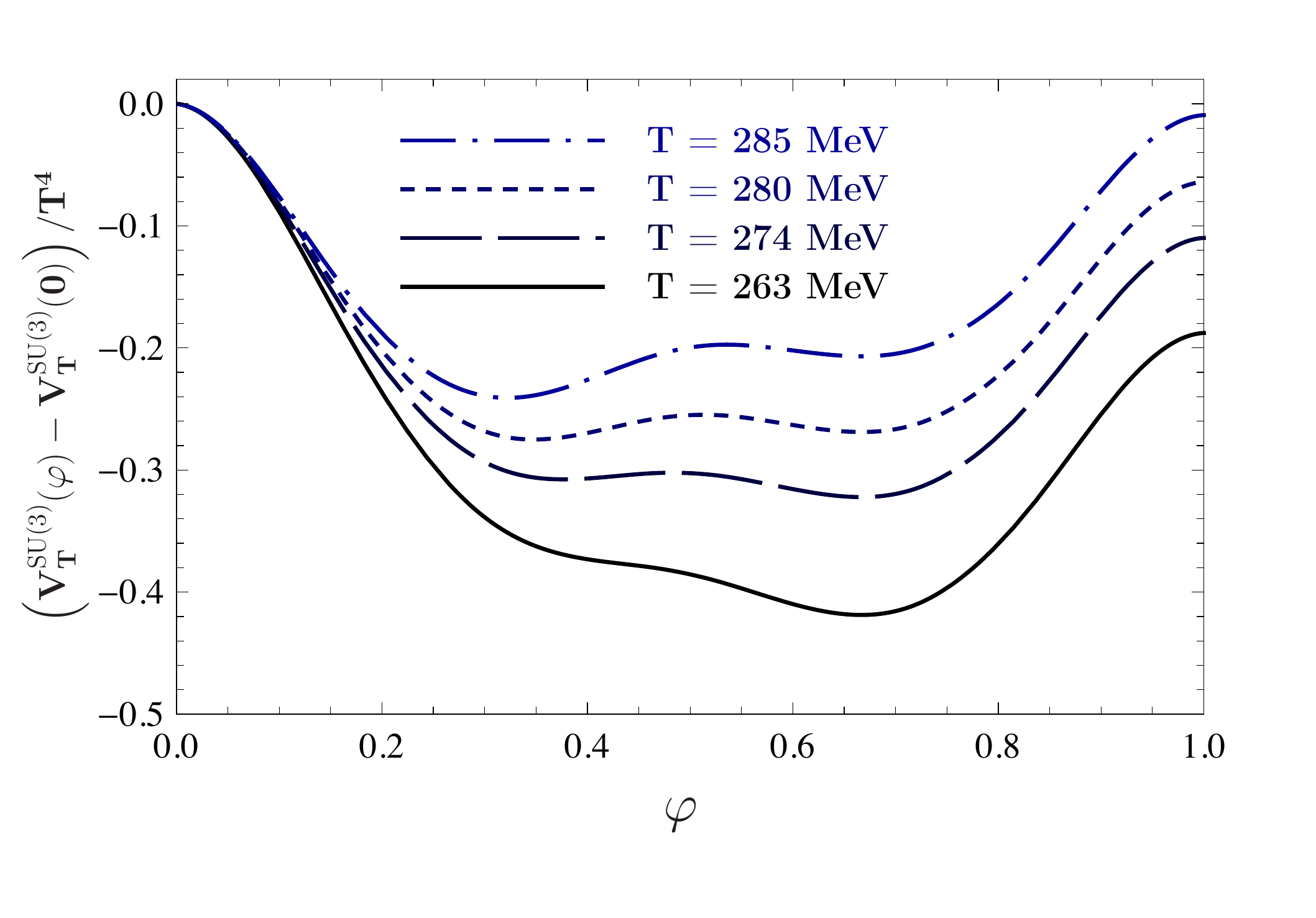}\label{fig:SU3_FRG_T}}
\caption[]{Polyakov loop potential for $SU(2)$ and $SU(3)$ obtained from the FRG.}
\label{fig:Pot_FRG_T}
\end{center}
\end{figure}

Here we present the results for $V[A_0]$ in the FRG, DSE and 2PI
approaches, which have previously been computed with the thermal
propagators obtained from the FRG formalism
\cite{Fister:2011uw,Fister:2011um, Fister:Diss}. The potential
determined from the flow equation \eq{eq:funflow} is depicted for
temperatures around the confinement-deconfinement phase transition for
$SU(2)$ in \Fig{fig:SU2_FRG_T}, and for $SU(3)$ in
\Fig{fig:SU3_FRG_T}.
The critical
temperatures $T_c$ are determined at
\begin{equation}\label{eq:FRG-Tc}
  T_{c}^{SU(2)}= 230\pm 23\, {\rm MeV}\ \ {\rm and}  \ 
  \ T_{c}^{SU(3)}= 275\pm 27\, {\rm MeV}\,.  
\end{equation} 
The absolute temperatures in \eq{eq:FRG-Tc} are set in comparison to
absolute scales of the lattice propagators \cite{Fischer:2010fx, Maas:2011ez,Maas:2011se} with a string tension
$\sigma= (420\,  {\rm MeV})^2\,$,
which leads to the dimensionless ratios of $T_c^{SU(2)}/\sqrt{\sigma}
\approx 0.548$ and $T_c^{SU(3)}/\sqrt{\sigma} \approx 0.655$,
respectively.  For $SU(3)$ our result is in quantitative agreement
with the lattice results, see e.g.\ \cite{Lucini:2012gg},
\begin{equation}\label{eq:lattice-Tc}
 T_{c,{\rm latt}}^{SU(2)}= 295\, {\rm MeV}\,, \ 
 T_{c,{\rm latt}}^{SU(3)}=270 \, {\rm MeV}\,,
\ {\rm thus,} \ \ \ 
\frac{T_{c,{\rm latt}}^{SU(2)}}{\sqrt{\sigma}} \approx0.709 \,,\ 
\frac{T_{c,{\rm latt}}^{SU(3)}}{\sqrt{\sigma}} \approx 0.646\,.
\end{equation}
Our relative normalisation is taken from the peak position of the
lattice propagators in \cite{Fischer:2010fx, Maas:2011ez,Maas:2011se}. This
normalisation has an error of approximately $10\%$, which sets the
errors.

For $SU(2)$ the critical temperature $T_c^{SU(2)}$ is
significantly lower than the lattice temperature in
\eq{eq:lattice-Tc}. This is linked to the missing back-reaction of the
fluctuations of $V[A_0]$ onto the propagators
\cite{Fister:2013bh}. Its inclusion in the Landau gauge flow leads to
a transition temperature of $T_c^{SU(2)}=300\pm 30\, {\rm MeV}$
\cite{Spallek}, in quantitative agreement with results in Polyakov
gauge \cite{Marhauser:2008fz} and from lattice gauge theory.

For the computation of $V[A_0]$ from its DSE, \eq{eq:DSEPolPot},
two-loop diagrams involving dressed vertices in the presence of a
background $A_0$ need to be determined. However, employing a
resummation scheme based on the 2PI-hierachy their contributions can
be minimised, which sets the optimal renormalisation point $\mu\approx
1\, {\rm GeV}$ \cite{Fister:2013bh}. As a consequence, it suffices to
evaluate a one-loop equation which does not involve dressed Green
functions other than the propagators.

\begin{figure}[t]
\begin{center}
\subfigure[Potential for $SU(2)$.]{\includegraphics[width=.48\textwidth]{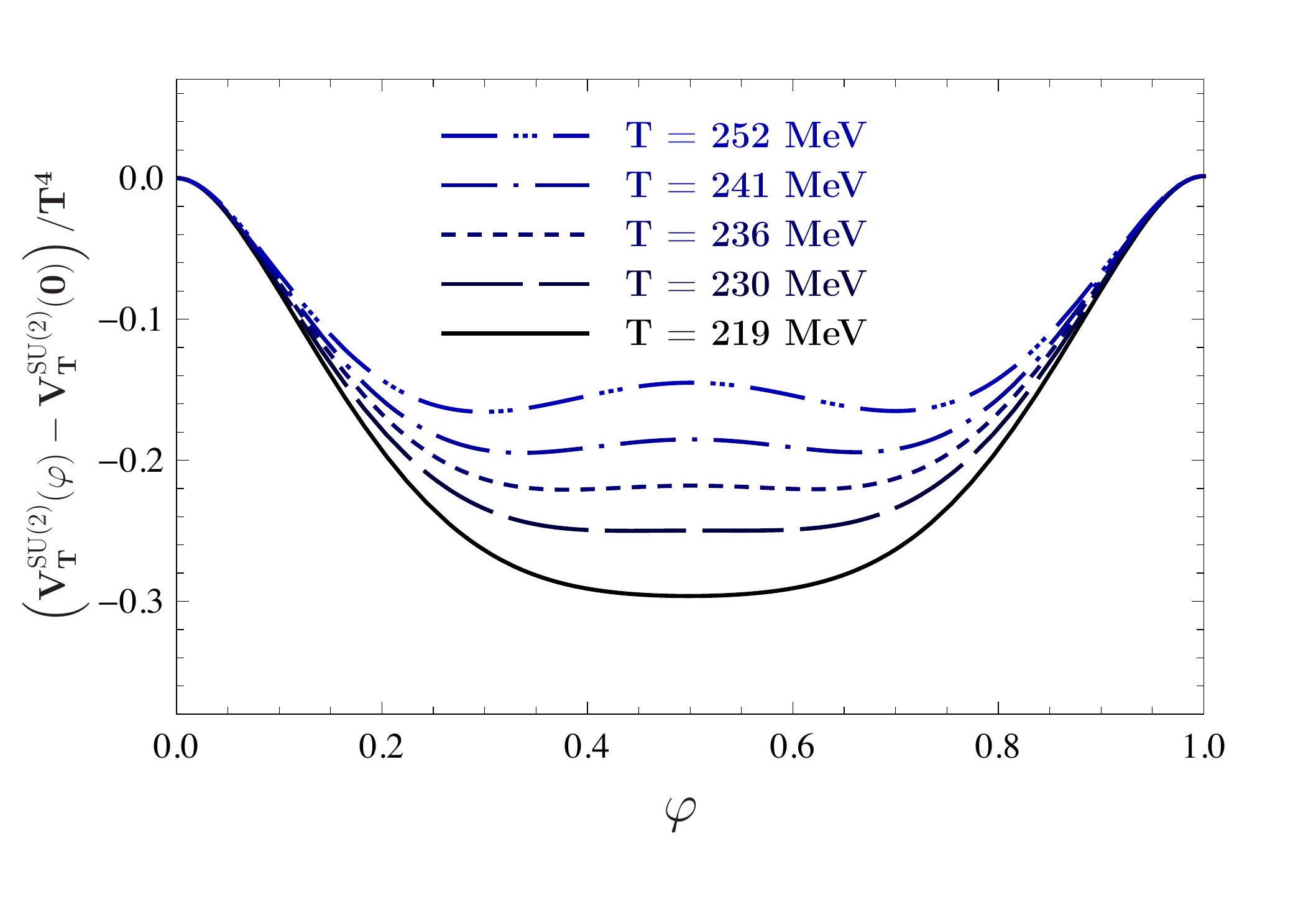} \label{fig:SU2_DSE_T}}
\subfigure[Potential for $SU(3)$.]{\includegraphics[width=.48\textwidth]{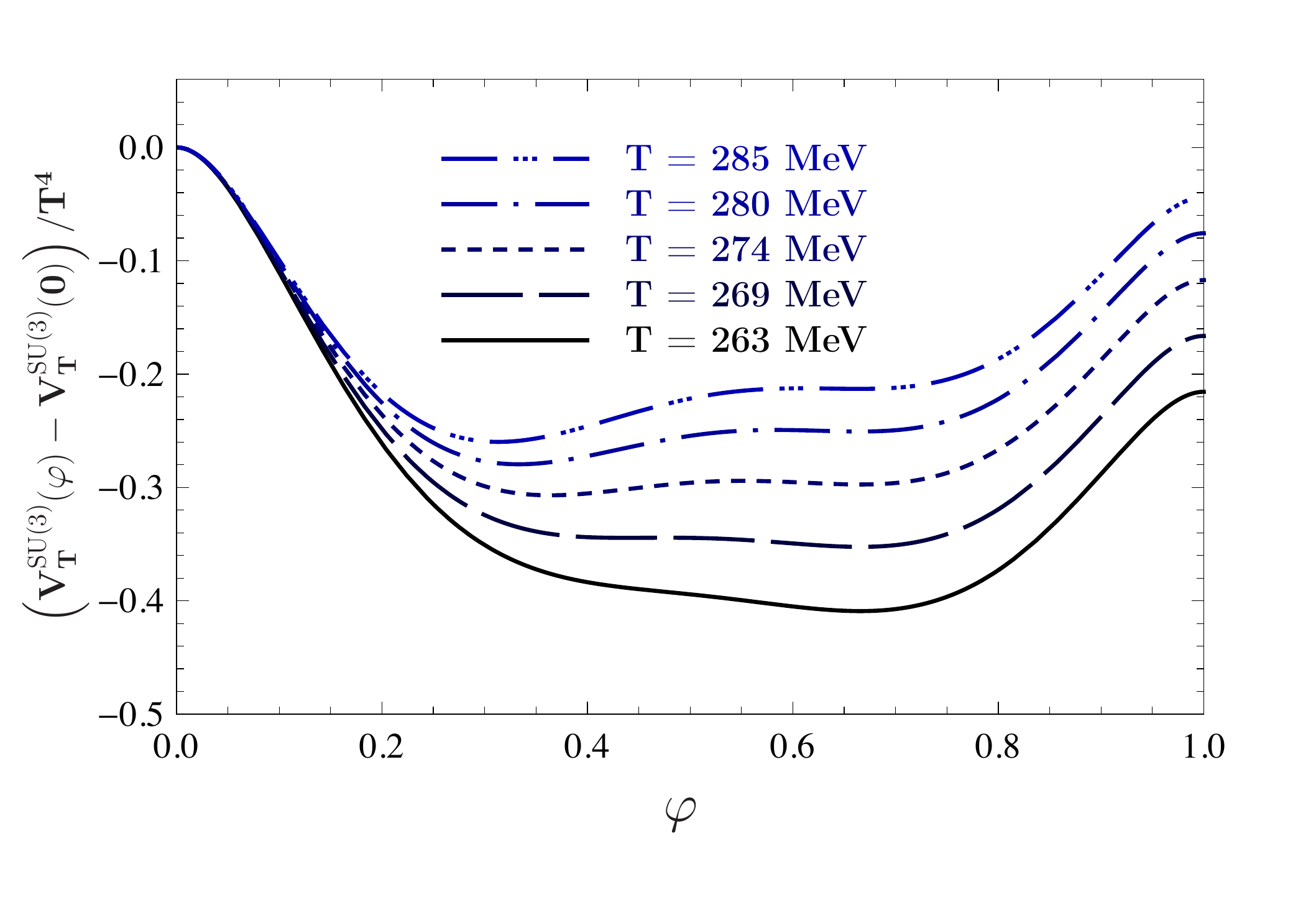} \label{fig:SU3_DSE_T}}
\caption[]{Polyakov loop potential for $SU(2)$ and $SU(3)$ from its DSE/2PI-representation within an optimised RG-scheme.}
\end{center}
\end{figure}

The one-loop diagrams in the DSE, \Fig{fig:DSE}, reduce to
\begin{eqnarray} \nonumber V'(\varphi)=&& \012 V'_{\rm
    Weiss}(\varphi)+\sumint_q \,2 \pi T\left(n+\varphi
  \right)\\[1ex]
  &&  \times \left[ z_a G_L\left(q+2 \pi T\varphi
    \right)+ 2\, z_a G_T\left(q
      +2 \pi T\varphi \right)\right.
\left.-2 z_c G_c\left(q+2 \pi T \varphi\right)\right] \,,
\label{eq:DSEpolpot}\end{eqnarray}
where $G_{L/T},\,G_{c}$ are the propagators of the
chromoelectric/chromomagnetic gluon and ghost normalised at
$z_{a/c}\!=\!Z_{a/c}(p^2=\mu^2)$ at zero temperature. The first term
originates in the trivial gauge mode. The evaluation of
\eq{eq:DSEpolpot} leads to the potential depicted in
\Fig{fig:SU2_DSE_T} for $SU(2)$ and \Fig{fig:SU3_DSE_T} for $SU(3)$.
From those the confinement-deconfinement critical temperatures are
determined at
\begin{equation}\label{eq:DSE+2PI-Tc}
  T_{c}^{SU(2)}= 235\pm 24\, {\rm MeV}\,, \  T_{c}^{SU(3)}= 275\pm 27\, {\rm MeV}\,, \ {\rm thus,} \quad
  \frac{T_{c}^{SU(2)}}{\sqrt{\sigma}}\approx 0.56\,, \   \frac{T_{c}^{SU(3)}}{\sqrt{\sigma}}\approx 0.655\,.
\end{equation}

\section{Conclusions}\label{sec:conclusions} 
We have studied the confinement-deconfinement phase transition of
static quarks with the Polyakov loop (effective) potential, which was
computed in the background field formalism in the Landau--deWitt gauge
\cite{Fister:2013bh}. This allowed for the use of Landau gauge
propagators at vanishing background, which had been obtained
previously in the framework of the FRG
\cite{Fister:2011um,Fister:2011uw, Fister:Diss}. The study here was
done within different non-perturbative functional continuum
approaches, i.e. using FRG-, DSE- and 2PI-representations of the
effective action.

In agreement with lattice gauge theory we find a second and first
order phase transition for $SU(2)$ and $SU(3)$, respectively.  The
critical temperatures in different methods agree quantitatively within
errors. These results are stable w.r.t.\ the choice of scaling or
decoupling type solutions for the propagators.  The critical
temperature for $SU(3)$ from functional methods agrees quantitatively
with lattice results. For $SU(2)$ the critical temperature from
functional methods is below the lattice temperature, which is due to
the missing back-reaction of the fluctuations of the Polyakov loop
potential onto the higher Green functions, in particular the
chromoelectric propagator.

The dependence of the Polyakov loop potential on the individual gluon
and ghost modes was exploited to derive a simple criterion for static
quark confinement. For infrared suppressed gluon but not suppressed
ghost propagators confinement is immanent at sufficiently low
temperatures, as the transversal gluons do not contribute and the
ghost modes overcompensate the deconfining longitudinal gluon mode.

In summary this work establishes the functional approach to
confinement with means of the Polyakov loop potential. It only
requires the background gauge propagators. Hence, it gives easy access
to structural questions about the confinement mechanism and to
explicit computations of the phase structure.

{\it Acknowledgments} --- This work is supported by the Helmholtz Alliance
HA216/EMMI and by ERC- AdG-290623. LF is supported by the
Science Foundation Ireland in respect of the Research Project
11-RFP.1-PHY3193 and by the Helmholtz Young Investigator Grant VH-NG-322.

\bibliographystyle{bibstyle}
\bibliography{Fister_ConfX}

\begin{thebibliography}{10}

\bibitem{Fister:2013bh}
L.~Fister and J.~M. Pawlowski,
\newblock (2013), 1301.4163.

\bibitem{Braun:2007bx}
J.~Braun, H.~Gies, and J.~M. Pawlowski,
\newblock Phys.Lett. {\bf B684}, 262 (2010), 0708.2413.

\bibitem{Marhauser:2008fz}
F.~Marhauser and J.~M. Pawlowski,
\newblock (2008), 0812.1144.

\bibitem{Braun:2009gm}
J.~Braun, L.~M. Haas, F.~Marhauser, and J.~M. Pawlowski,
\newblock Phys.Rev.Lett. {\bf 106}, 022002 (2011), 0908.0008.

\bibitem{Pawlowski:2010ht}
J.~M. Pawlowski,
\newblock AIP Conf.Proc. {\bf 1343}, 75 (2011), 1012.5075.

\bibitem{Braun:2010cy}
J.~Braun, A.~Eichhorn, H.~Gies, and J.~M. Pawlowski,
\newblock Eur.Phys.J. {\bf C70}, 689 (2010), 1007.2619.

\bibitem{Reinhardt:2012qe}
H.~Reinhardt and J.~Heffner,
\newblock Phys.Lett. {\bf B718}, 672 (2012), 1210.1742.

\bibitem{Diakonov:2012dx}
D.~Diakonov, C.~Gattringer, and H.-P. Schadler,
\newblock JHEP {\bf 1208}, 128 (2012), 1205.4768.

\bibitem{Greensite:2012dy}
J.~Greensite,
\newblock Phys.Rev. {\bf D86}, 114507 (2012), 1209.5697.

\bibitem{Fukushima:2012qa}
K.~Fukushima and K.~Kashiwa,
\newblock (2012), 1206.0685.

\bibitem{Kashiwa:2012td}
K.~Kashiwa and Y.~Maezawa,
\newblock (2012), 1212.2184.

\bibitem{Fister:2011uw}
L.~Fister and J.~M. Pawlowski,
\newblock (2011), 1112.5440.

\bibitem{Fister:2011um}
L.~Fister and J.~M. Pawlowski,
\newblock (2011), 1112.5429.

\bibitem{Fister:Diss}
L.~Fister,
\newblock PhD thesis, Heidelberg University  (2012).

\bibitem{Fischer:2006vf}
C.~S. Fischer and J.~M. Pawlowski,
\newblock Phys. Rev. {\bf D75}, 025012 (2007), hep-th/0609009.

\bibitem{Fischer:2009tn}
C.~S. Fischer and J.~M. Pawlowski,
\newblock Phys. Rev. {\bf D80}, 025023 (2009), 0903.2193.

\bibitem{Fischer:2008uz}
C.~S. Fischer, A.~Maas, and J.~M. Pawlowski,
\newblock Annals Phys. {\bf 324}, 2408 (2009), 0810.1987.

\bibitem{Fister:2010ah}
L.~Fister,
\newblock (2010), 1002.1649.

\bibitem{Fister:2010yw}
L.~Fister, R.~Alkofer, and K.~Schwenzer,
\newblock Phys.Lett. {\bf B688}, 237 (2010), 1003.1668.

\bibitem{Alkofer:2010tq}
R.~Alkofer, L.~Fister, A.~Maas, and V.~Macher,
\newblock AIP Conf.Proc. {\bf 1343}, 179 (2011), 1011.5831.

\bibitem{Macher:2011ys}
V.~Macher, A.~Maas, and R.~Alkofer,
\newblock Int.J.Mod.Phys. {\bf A27}, 1250098 (2012), 1106.5381.

\bibitem{Braun:2011fw}
J.~Braun and A.~Janot,
\newblock Phys.Rev. {\bf D84}, 114022 (2011), 1102.4841.

\bibitem{Braun:2012zq}
J.~Braun and T.~K. Herbst,
\newblock (2012), 1205.0779.

\bibitem{Fischer:2010fx}
C.~S. Fischer, A.~Maas, and J.~A. M{\"u}ller,
\newblock Eur. Phys. J. {\bf C68}, 165 (2010), 1003.1960.

\bibitem{Maas:2011ez}
A.~Maas, J.~M. Pawlowski, L.~von Smekal, and D.~Spielmann,
\newblock Phys.Rev. {\bf D85}, 034037 (2012), 1110.6340.

\bibitem{Maas:2011se}
A.~Maas,
\newblock Phys. Rep. in press  (2013), 1106.3942.

\bibitem{Lucini:2012gg}
B.~Lucini and M.~Panero,
\newblock (2012), 1210.4997.

\bibitem{Spallek}
F.~Spallek,
\newblock Diploma thesis, Heidelberg University  (2010).

\end{thebibliography}

\end{document}